\documentclass[prc,aps,twocolumn,amssymb,aps,nofootinbib,floatfix]{revtex4}
\usepackage[ansinew]{inputenc}
\usepackage[T1]{fontenc} 
\usepackage{graphicx}
\usepackage{epsfig}

\newcommand{\GeV}{\hbox{ GeV}}

\newcommand{\fm}{\hbox{ fm}}

\begin{document}

\title{Negative Elliptic Flow of $J/\psi$'s: A Qualitative Signature for Charm Collectivity at RHIC}

\author{Daniel Krieg and Marcus Bleicher}
\affiliation{Institut f\"ur Theoretische Physik, Johann Wolfgang Goethe-Universit\"at, 
Max-von-Laue-Str.~1, 60438 Frankfurt am Main, Germany}
\date{September 4, 2007}

\begin{abstract}
We discuss one of the most prominent features of the very recent preliminary elliptic flow data of $J/\psi$ meson 
from the PHENIX collaboration \cite{Silvestre:2008tw}. Even within the the rather large error bars of the measured data 
a negative elliptic flow parameter ($v_2$) for $J/\psi$ in the range of $p_T=0.5-2.5 \GeV/c$ is visible. 
We argue that this negative elliptic flow at intermediate $p_T$ is a clear and qualitative signature for the collectivity
of charm quarks produced in nucleus-nucleus reactions at RHIC. Within a parton recombination approach we show
that a negative elliptic flow puts a lower limit on the collective transverse velocity of heavy quarks. The numerical
value of the transverse flow velocity $\beta_T$ for charm quarks that is necessary to reproduce the 
data is $\beta_T(charm)\sim 0.55-0.6c$ and therefore compatible with the flow of light quarks.
\end{abstract}

\maketitle

The main goal of the current and past heavy ion programs is the search for a new state of matter
called the Quark-Gluon-Plasma (QGP) \cite{Bass:1998vz}. Major breakthroughs for the 
potential discovery \cite{Adcox:2004mh,Adams:2005dq} of this new state of matter were the observation of constituent
quark number scaling of the elliptic flow $v_2^{\rm hadron} (p_T^{\rm hadron}) = n_q v_2^{q} (p_T^{\rm hadron}/n_q)$, 
with $n_q$ being the number of constituent quarks in the respective hadron as well as the observation of jet quenching
at intermediate transverse momenta \cite{Wang:2005yc,Vitev:2005he,Armesto:2005iq}.
Together with the 'standard' hydrodynamical interpretation this implies a rapid thermalization and a strong 
collective flow of the QCD matter created at RHIC. However, open questions remain: how can one obtain a 
consistent description of the high $p_T$ suppression and 
the elliptic flow of heavy flavour quarks and hadrons. I.e. is the collectivity at RHIC restricted to light 
quarks (up, down, strange) or do even charm (bottom) quarks participate in the collective expansion of 
the partonic system and reach local kinetic equilibrium?

Previously, it was assumed that local equilibrium of (heavy) quarks could not be achieved within pQCD transport 
simulations. In fact, older studies
\cite{Molnar:2001ux} based on a  parton cascade dynamics restricted to $2\leftrightarrow 2$ parton interactions 
seemed to indicate that the opacity needed to achieve local equilibrium
would be at least an order of magnitude higher than pQCD estimates. However, recent state-of-the-art parton cascade 
calculations (including $2\leftrightarrow 3$ parton interactions) have clearly shown that pQCD cross sections 
are sufficient to reach local (gluon) equilibrium and allow to describe the measured 
elliptic flow data \cite{Xu:2007aa,Xu:2007ns,Xu:2007jv,El:2007vg}.

The aim of the present letter is to investigate whether also the charm quark does 
locally equilibrate and therefore follows the flow of the light quarks.
Here we will focus on the $J/\psi$ because it reflects the momentum distribution of the
charm quarks directly, in addition first experimental data on the $J/\psi$ elliptic flow just became available. 
We will show that the recently measured negative elliptic flow of $J/\psi$'s provides a unique {\it lower} bound on
the charm quark's collective velocity.

Under the assumption of local equilibration of light quarks a hydrodynamic parametrization 
of the freeze-out hyper-surface to parametrize the quark emission function, namely the blast-wave model, can be employed. 
For the charm quarks, the same emission function is used, however, with the transverse collective velocity 
as a free parameter to be determined by the preliminary PHENIX data.
To calculate $J/\psi$'s from the charm quark emission function, we apply the well known parton 
recombination approach \cite{Zimanyi:1999py,Fries:2003kq,Krieg:2007sx}. Details (like the exact form of the 
freeze-out hyper-surface) of the specific approach employed here can be found in \cite{Fries:2003kq,Krieg:2007sx}. Different from there we used a linear increasing transverse flow rapidity instead of a constant one, but the mean value has been preserved.

Here we summarize the most important features: In a coalescence process the quarks contribute equally to 
the hadrons momentum, so it inherits its azimuthal asymmetry directly from its constituents. Therefore in 
recombination the elliptic flow of $J/\psi$'s emerges directly from a negative $v_2$ of the charm quark. 
To incorporate the asymmetry, the transverse expansion rapidity $\eta_T$ depend on the azimuthal angle $\phi$ and the radial coordinate $\rho=\frac{r}{R}$ as
\begin{equation}
\eta_T(\phi,\rho) = \eta_T^0 \cdot \frac{3}{2}\rho \left(1+\varepsilon f(p_T) \cos(2\phi)\right)
\end{equation}
with the eccentricity $\varepsilon$ and $f(p_T)=1/\left(1+(p_T/p_0)^2\right)$ to model the damping at high $p_T$. 
With the factor $\frac{3}{2}$ we recover $\eta_T^0$ as the mean transverse rapidity after integrating over $\rho$.

By applying the definition of the elliptic flow one obtains \cite{Fries:2003kq}
\begin{equation}
v_2^q(p_T) = \frac{\int \cos(2\phi) I_2 \left[a(\phi,\rho)\right] K_1 \left[b(\phi,\rho)\right] \, d\phi\,  \rho\, d\rho}
{\int I_0 \left[a(\phi,\rho)\right] K_1\left[b(\phi,\rho)\right] \, d\phi\, \rho\, d\rho}
\end{equation}
with $a(\phi,\rho) = p_T \sinh(\eta_T(\phi,\rho)) /T$, $b(\phi,\rho) = m_T \cosh(\eta_T(\phi,\rho)) /T$
and the modified Bessel functions $I_n$ and $K_n$. For a more general hydrodynamical hypersurface one could assume a dependence of freeze-out time $\tau$ on the radial coordinate $\rho$. This would lead to additional terms involving $\frac{\partial \tau}{\partial \rho}$ and Bessel functions of other order. We have checked that the modifications are only minor and therefore neglect the contributions in this letter for brevity.

\begin{figure}[hb]
 \centering
\includegraphics{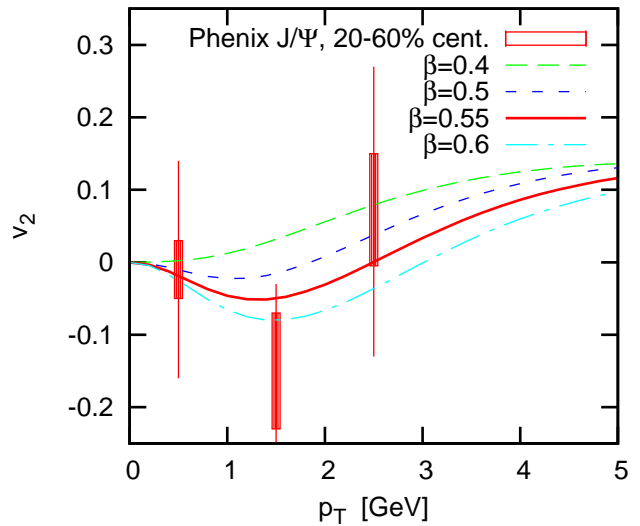}
 \caption{Elliptic flow ($v_2$) of $J/\psi$'s for $b=9\fm$ for different mean transverse expansion velocities (lines) 
compared to preliminary data from PHENIX collaboration \cite{Silvestre:2008tw}. While the $v_2$ of $J/\psi$'s is smaller than for light hadrons, 
the mean transverse velocity for the best-fit case ($\beta=0.55c$ for charm quarks) is the same as 
for light quarks.}
 \label{plt:jpsi_flow}
\end{figure}

Let us investigate the elliptic flow of the $J/\psi$ at midrapidity as a function of the transverse momentum for various
transverse flow velocities as shown in Fig. \ref{plt:jpsi_flow}. The lines from top to bottom indicate calculations with a charm quark mass $m_c = 1.5 \GeV/c^2$ for different mean expansion velocities $\tanh(\eta_T^0)=\beta_T=0.4c,0.5c,0.55c,0.6c$, the data by the PHENIX collaboration are
shown as symbols with error bars indicating a negative elliptic flow for $J/\psi$'s at intermediate transverse momenta. 
The calculation shows that with increasing transverse flow a negative $v_2$ at low $p_T$ (above $p_T\sim 2.5$~GeV,the elliptic flow values turn positive again) develops for the $J/\psi$, posing a lower
bound of $\beta_T\geq 0.5c$ for the charm quarks flow. The best fit to the data is obtained with a mean charm flow velocity
of $\beta_T= 0.55c$ equal to the light quark flow velocity extracted from previous fits within the same model.

\begin{figure}
 \centering
\includegraphics{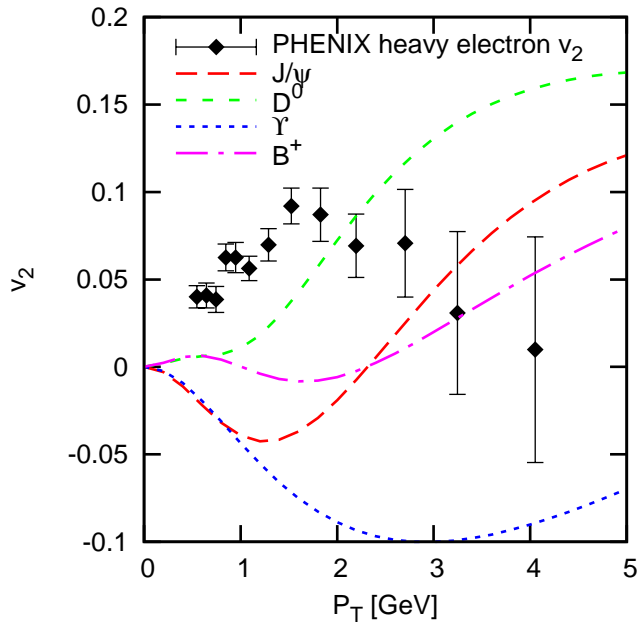}
 \caption{Comparison of elliptic flow ($v_2$) for $J/\psi$, $D^0$, $\Upsilon$ and $B^+$ at $b=9\fm$ with $\beta=0.55c$ to data of non-photonic electrons from PHENIX collaboration \cite{Adare:2006nq}.}
 \label{plt:compare_flow}
\end{figure}

\begin{figure}
 \centering
\includegraphics{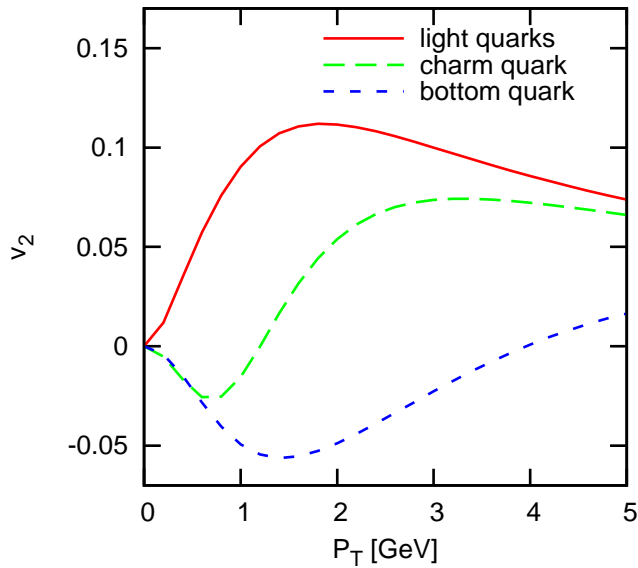}
 \caption{Elliptic flow ($v_2$) of light, charm and bottom quarks at $b=9\fm$ with $\beta=0.55c$.}
 \label{plt:quark_flow}
\end{figure}

In Fig.~\ref{plt:compare_flow} we use $\beta_T= 0.55c$ and compare the elliptic flow to other heavy mesons and Fig.~\ref{plt:quark_flow} shows the same for the quarks. The value for $\Upsilon$, with a bottom quark mass of $m_b = 4.7 \GeV/c^2$, is negativ in the whole range of applicability. In contrast to $J/\psi$ , the $v_2$ of $D^0$ stays positiv. This is due to the positive light quark $v_2$, which competes with the negative one for the charm quark, and results in nearly zero elliptic flow at low $p_T$. While the $B^+$ meson follows the $D^0$ flow for $p_T<1\GeV$, it is much more suppressed at higher $p_T$ due to the strong negative flow of the bottom quark and approximately zero up to $p_T\sim2.5\GeV$.

Data on $D$-meson elliptic flow is not yet available. When comparing it to the non-photonic electron $v_2$, our calculations fail to predict the data \cite{Sakai:2007zzb}. These two observables have been predicted to be similiar \cite{Greco:2003vf}, since the non-photonic electrons are mainly from , $D$-meson decays, but with a small contribution of $B$-meson decays. But the electron elliptic flow is no straightforward probe for the $D^0$ $v_2$. Since the electron is not the only decay product, the decay kinematics might smear out the resulting elliptic flow of the electrons.
At low $p_T$, the increase of the $D^0$ flow is similiar to the electron data, but shifted to higher transverse momenta. Above $p_T=2\GeV$ the electron $v_2$ starts to decrease which might be due to contributions from the $B$-mesons or an early onset of the fragmentation regime.

Direct measurements on the elliptic flow of heavy mesons will be available in the near future with the heavy-flavor tracker for STAR, which will allow a better analysis. Therefore the presented results are based only the $J/\psi$ elliptic flow data.

These results provide strong evidence for a substantial collectivity and transverse expansion of the charm quarks
in nucleus-nucleus reactions at RHIC. Due to the large error bars this has to be verified when more precise data is available. Note that our present findings are different from previous approaches that assume
incomplete thermalization of the charm \cite{Greco:2003vf,Yan:2006ve,Ravagli:2007xx,Linnyk:2008uf}. We also verified our findings within a boltzmann approach to coalescence \cite{Ravagli:2007xx} using our parametrizations and received similar results.

One should also note that the observation of negative elliptic flow of heavy particles is well known
in the literature (even if not conclusively observed experimentally up to now). It appears due to an interplay between
transverse expansion and particle mass, the more flow and the heavier the particle the more negative values does the
elliptic flow reach. E.g., negative values of the elliptic flow parameter for heavy hadrons has also been found 
in previous exploratory studies and seem to be a general feature of the blast-wave like flow profile
at high transverse velocities \cite{Voloshin:1996nv,Huovinen:2001cy,Voloshin:2002ii,Retiere:2003kf,Pratt:2004zq}.
It reflects the depletion of the low $p_T$ particle abundance, when the source elements are highly boosted
in the transverse direction. The difference to the present study is that here, $v_2$ is already negative on 
the quark level. Negative elliptic flow values will even be 
encountered for light quarks at asymptotically high bombarding energies as discussed in \cite{Krieg:2007sx}.
One might argue that this is an artefact of the blast-wave peak and will not survive in more realistic
calculations, however also transport model calculations show slightly negative $v_2$ values for heavy particles 
at low transverse momenta \cite{Bleicher:2000sx}.

In conclusion, we have shown that the recent preliminary PHENIX data exhibiting a negative elliptic flow at low $p_T$ can be explained within a parton recombination approach using a blast-wave like parametrization. We point out that studying $v_2(p_T)$ from $J/\psi$ offers the possibility to put a lower limit on the charm quark transverse velocity. From the present quantitative analysis we expect the transverse velocity of charm quarks to be above $\beta_T\geq 0.4c$. Within the limits of the present model the
best description of the data is obtain for a charm transverse velocity equal to the light quark velocity 
of $\beta_T=0.55c$. So if more precise data will still support the negative $v_2$, we conclude from this observation that charm quarks reach a substantial amount of local 
kinetic equilibration.

\section*{Acknowledgements}

This work was supported by Gesellschaft f\"ur Schwerionenforschung, Darmstadt (GSI) and by the LOEWE initiative 
of the State of Hessen through the Helmholtz International Center for FAIR.

\end{document}